\documentclass{elsart3}

\usepackage{epsfig}
\usepackage{amssymb}

\begin{document}

\begin{frontmatter}



\title{IS DEMAGNETIZATION AN EFFICIENT OPTIMIZATION METHOD?}


\author[1]{S. Zapperi}
\author[1]{F. Colaiori}
\author[1,2]{L. Dante}
\author[3]{V. Basso}
\author[3]{G. Durin}
\author[3]{A. Magni}
\author[1,4]{M. J. Alava}
\address[1]{INFM-SMC, Dipartimento di Fisica,
Universit\`a "La Sapienza", P.le A. Moro 2
00185 Roma, Italy}
\address[2]{CNR Istituto di Acustica ``O. M. Corbino'', via 
del fosso del Cavaliere 100, 00133 Roma, Italy}
\address[3]{Istituto Elettrotecnico Nazionale Galileo Ferraris, 
strada delle Cacce 91, I-10135 Torino, Italy}
\address[4]{Helsinki University of Technology, Laboratory of Physics,  
HUT-02105 Finland}
\begin{abstract}
Demagnetization, commonly employed to study ferromagnets,
has been proposed as the basis for an optimization tool, a method to
find the ground state of a disordered system. 
Here we present a detailed comparison between
the ground state  and the demagnetized state  
in the random field Ising model, combing exact results in
$d=1$  and numerical solutions in $d=3$.  We show that there are
important differences between the two states that persist 
in the thermodynamic limit and thus conclude that AC demagnetization is not 
an efficient optimization method.
\end{abstract}

\end{frontmatter}


Disordered systems are widely studied both 
for their conceptual importance and because the presence of randomness 
provides prototypical examples of complex optimization problems 
\cite{ALA-01}. 
A disordered system can be non-trivial even at zero temperature due to
the presence of a complex energy landscape. The properties of the
ground-state (GS) are often difficult to determine analytically, and
exact numerical evaluation becomes computationally prohibitive for large
systems. Thus one is lead to construct approximate schemes, typically
based on a non--equilibrium dynamics to find low energy states. In this
respect a  recently proposed method is hysteretic optimization \cite{ZAR-02}.
Its basis is an analogy to a ferromagnetic demagnetization procedure:
 an external oscillating field with decreasing amplitude and 
low frequency is applied to the
system.  In ferromagnetic materials, one obtains at zero field
after this procedure the demagnetized state (DS), which is used as a
reference state for material characterization.

Here we analyze the differences between the GS and the DS in the 
ferromagnetic random field Ising model (RFIM), which has been extensively
studied in literature as a paradigmatic example of disordered system
\cite{NAT-00}.
In the RFIM, a spin $s_i = \pm 1$ is assigned to each
 site $i$ of a $d-$dimensional
lattice. The spins are coupled to their nearest-neighbors 
spins by a ferromagnetic interaction of strength $J$ and 
to  the external field $H$. In addition, to each site of
the lattice it is associated a random field $h_i$ taken from a Gaussian 
probability $\rho(h)=\exp(-h^2/2R^2)/\sqrt{2\pi}R$), with variance $R$.
The Hamiltonian thus reads
\begin{equation} 
F = -\sum_{\langle i,j \rangle}Js_i s_j -
\sum_i(H+h_i)s_i,\label{eq:rfim}
\end{equation}
where the first sum is restricted to nearest-neighbors pairs.
The $T=0$ equilibrium
problem amounts to find the minimum  of $F$ for a given
realization of the random-fields (i.e. the GS). This can be
achieved for in a polynomial CPU-time \cite{ALA-01}, with exact combinatorial algorithms. 
The one dimensional case can instead be solved exactly \cite{BRU-84}.

For the out of equilibrium case, we consider the 
dynamics proposed  in Refs.~\cite{SET-93}: 
at each time step the spins align with the local field
$s_i = \mbox{sign}(J\sum_j s_j  + h_i +H)$,
until a metastable state is reached. This dynamics can be used 
to obtain the hysteresis loop. The system is started from a 
state with all the spin down $s_i=-1$ and then $H$ is ramped
slowly from  $H \to -\infty$  to $H \to \infty$. 

The main hysteresis loop selects a series of metastable states,
which in principle are not particularly close to the ground state.
To obtain low energy states, we perform a demagnetization procedure:
the external field is changed through a nested succession $H =
H_0 \to H_1 \to H_2 \to ..... H_n...\to 0$, with  
$H_{2n}>H_{2n+2}>0$, $ H_{2n-1}<H_{2n+1}<0$ and
$dH
\equiv H_{2n}-H_{2n+2}\to 0$. 
In $d=3$ simulations we perform an approximate 
demagnetization using $dH=10^{-3}$,
while in $d=1$ we obtain the exact solution in the limit 
$dH/dt\to 0$ \cite{DAN-02}.

\begin{figure}[ht]
\centerline{\epsfig{file=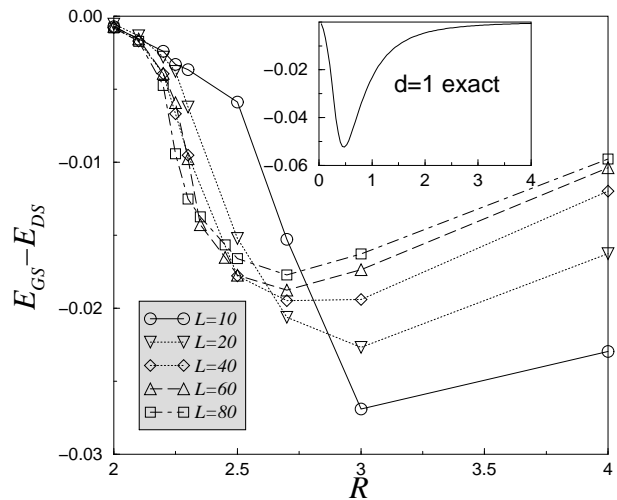,width=8cm,clip=!}}
\caption{The energy difference  between  
the GS and the DS in $d=3$ as a function of disorder
for different system sizes. In the inset we show the same
curve calculated exactly in $d=1$.}
\label{figde}
\end{figure}

In Fig.~\ref{figde} we plot the energy difference between
the GS and the DS in $d=3$ as a function of disorder
for different system sizes $L$. While the curves clearly change
with the system size, there is no indication that in the thermodynamic
limit $L\to \infty$ the two energies converge to the same value.
This is confirmed by the exact result in $d=1$, reported in the inset,
which is obtained directly in the thermodynamic limit. Also in this
case there is a region in which the energies of the GS and the DS
differ substantially. The energy differences
are small when disorder is large, since almost each spin 
trivially aligns with its local field. I\cite{NAT-00,SET-93}.
For low disorder, deep in the ferromagnetic state, which is only present 
in $d=3$ \cite{NAT-00,SET-93}, the differences between GS and DS are again
small,  reflecting the fact that almost all spins point 
in one direction. 

In conclusions, our analysis indicate that demagnetization is not
an efficient optimization tool, apart from the cases in which the
DS and GS are trivial. We have performed a detailed analysis of the
domain structures in the DS and the GS and found that the domain
 structures are not easily related by local spin flips. This cast
some doubt on the wide applicability of optimization algorithms based
on demagnetization.

\end{document}